# Cross-domain recommender system using Generalized Canonical Correlation Analysis

S. M. Hashemi, and M. Rahmati

**Abstract:**

Recommender systems provide personalized recommendations to the users from a large number of possible options in online stores. Matrix factorization is a well-known and accurate collaborative filtering approach for recommender system, which suffers from cold-start problem for new users and items. Whenever a new user participate with the system there is not enough interactions with the system, therefore there are not enough ratings in the user-item matrix to learn the matrix factorization model. Using auxiliary data such as user's demographic, ratings and reviews in relevant domains, is an effective solution to reduce the new user problem. In this paper, we used data of users from other domains and build a common space to represent the latent factors of users from different domains. In this representation we proposed an iterative method which applied MAX-VAR generalized canonical correlation analysis (GCCA) on user's latent factors learned from matrix factorization on each domain. Also, to improve the capability of GCCA to learn latent factors for new users, we propose generalized canonical correlation analysis by inverse sum of selection matrices (GCCA-ISSM) approach, which provides better recommendations in cold-start scenarios. The proposed approach is extended using content-based features from topic modeling extracted from user's reviews. We demonstrate the accuracy and effectiveness of the proposed approaches on cross-domain ratings predictions using comprehensive experiments on Amazon and MovieLens datasets.

## 1 INTRODUCTION

With the development of Internet in recent years, online marketing websites, online music and movie stream services play an important role in e-commerce. At the beginning, these websites were just a database of different products, which provided the ability to search products in different categories. Gradually, and as the number of products increased, users faced more options to choose from these websites. For example, in an online market, as the number of products increases, finding a movie among the large number of existing movies becomes difficult and users encounter information overload. In order to solve this problem, new features such as personal search and product recommendations based on the user's interest are added to these sites, which not only improve user experience and increase user satisfaction, but also result in more purchases and increase profits of marketing websites.

In order to suggest a personalized recommendation to the user, demographic information, purchase history, previous item's ratings, reviews, comments and etc. may be considered. The intelligent system employing this information to offer an item to the user for purchase is called recommender system. Some of the most important applications of recommender system are offering products in online marketing websites, recommending movies[1] and music[2],[3] in media stream websites, and recommending personalized news from news websites [4]. Sparsity of user-item matrix, cold-start, low variety, scalability and using contextual information are some of common challenges in recommender systems[5]; among these challenges the cold-start problem is the most important one. In cold-start, since there is few ratings available for a user or an item, an appropriate recommendation cannot be offered. The cold-start challenge can be divided into two separate problems, new item, and new user problems [6]. New item problem occurs when a new item is registered in the system while no ratings or reviews has been entered for it; this problem is less challenging compared to new user problem, since new items can be suggested to the user using other methods like advertisements. New user problem is the most difficult challenges in recommender systems [5], since new users have not entered any ratings in the recommender system, the system cannot present any personalized recommendations. In order to solve the cold-start problem, two approaches are proposed. First, motivating the users to enter ratings and reviews in the system, second, using cross-domain recommendation algorithms which utilize existing information in other domains such as ratings, reviews, tags, demographics, social network relationships and etc. to improve the recommendation quality in the target domain.

[S].M.Hashemi is Ph.D. candidate in Computer Engineering and Information Technology, Amirkabir university of Technology (Tehran Polytechnic), Tehran, Iran. (e-mail:mohammadhme@aut.ac.ir)

M.Rahmati is with the Computer Engineering and Information Technology Department, Amirkabir University of Technology (Tehran Polytechnic), Tehran, Iran. (e-mail:rahmati@aut.ac.ir)

Many of the traditional recommender systems are based on single domain collaborative filtering. While some recommender systems like online marketing websites include different domains such as movies, books, music, videogames, clothing and etc. In such systems, knowledge aggregation or knowledge transfer from different domains can be used instead of constructing a recommendation model for each domain separately. For example, by transferring the genre of user's favorite movie to the book domain, books with similar genre can be recommended to the user. Such recommender systems which employ different domains to offer recommendations are called cross-domain recommender system. Cantador et al. [7] published a comprehensive survey on the cross-domain recommender systems approaches proposed in recent years. They specified the most important objectives of cross-domain recommender systems as solving cold-start challenge, increasing accuracy and variety.

In this paper, we proposed a new approach for cross-domain recommender system, which uses knowledge from auxiliary domains to solve the cold-start problem. Our approach is based on collaborative filtering and uses a customized version of generalized canonical correlation analysis (GCCA) for transferring knowledge from auxiliary domains to the target domain. To the best of our knowledge, this study is the first work to use GCCA in the recommender systems. Sahebi et al. [8] has used CCA for reconstructing user-item matrix in the target domain. Since they applied CCA on user-item matrix, it needs more memory to compute canonical variables and it can use user-item matrix of only one domain as auxiliary data.

The main contributions of this paper includes: (1) We proposed a novel iterative method using GCCA and matrix factorization for cross-domain ratings prediction. This is the first work that uses GCCA in the recommender system. (2) We propose a new constraint in GCCA to improve its ability to learn latent features for new users. (3) The proposed algorithm can use multiple auxiliary domains and other information like demographic and content-based features as auxiliary data. 4) We perform comprehensive experiments to explore weaknesses and strengths of the proposed algorithm.

The notations used through the paper are listed in Table 1. The rest of this paper is organized as follows: In the next section, we present an overview of recent works in recommender systems. Next, we formulate the cross-domain recommendation problem and present the proposed method. In Section 4 a simulation study on the proposed method is presented. In Section 5 we analyze the accuracy and performance of the proposed method on popular recommender system datasets and compare it with the baseline approaches. The conclusion is presented in Section 6.

TABLE 1
NOTATIONS

| Symbol | Description |
|--------|-------------|
| $\mathbb{U}$ | Set of all users |
| $\mathbb{V}$ | Set of all items |
| $\mathbb{U}_j$ | Set of users rated item $j$ |
| $\mathbb{V}_i$ | Set of items rate by user $i$ |
| $R^{(d)}$ | User-item matrix in domain $d$ |
| $r_{i,j}^{(d)}$ | Rating of user $i$ to item $j$ in domain $d$ |
| $n$ | Number of domains |
| $\mathbb{D}$ | Set of all domains |
| $L$ | Number of users |
| $U^{(d)}$ | Latent factor matrix of users in domain $d$ |
| $V^{(d)}$ | Latent factor matrix of items in domain $d$ |
| $k_d$ | Number of user's latent factors in domain $d$ |
| $K^{(d)}$ | Selection matrix in domain $d$ |
| $X^{(d)}$ | Observation matrix in view $d$ |
| $G$ | Group configuration of GCCA |
| $W^{(d)}$ | Canonical components matrix in view $d$ |
| $\Omega$ | Set of user-item pairs that has ratings in user-item matrix |

## 2 RELATED WORKS

In [5] the recommender system approaches are classified into 4 classes, including content-based filtering, demographic filtering, collaborative filtering and hybrid methods. In content-based filtering, features are extracted from item's content including text, image, audio and etc. Then items with features similar to the user's previous selection are recommended. In demographic filtering it is assumed the users with similar personal characteristics (sex, race, religion and etc.) have similar interests and would make similar choices [9]. In collaborative filtering methods [10], previous ratings of users are used to find similarity between users or items. It is assumed that users who have made similar choices in the past, would also make similar choices in the future. Therefore, the choices of a user can be suggested to other similar users. Hybrid methods uses combinations of collaborative filtering and demographic filtering [11], or collaborative filtering and content-based methods [12], [13] to improve the recommendations.

Collaborative filtering is the most common approach in recommender system and can be divided into memory-based and model-based methods. Memory-based methods focus on the relationship between items or users. These methods usually employ a similarity measure like cosine similarity or Pearson correlation and compute similarity between rows and columns of user-item matrix to find similar users or items [14]. The advantage of memory-based methods is that their results are always up-to-date, and its disadvantages are scalability problem and finding an appropriate similarity function [15], [16]. In recent years, model-based methods like Matrix factorization(MF) or SVD[17], SVD++[18] and etc. have attracted more attention due to their high accuracy and scalability. In matrix factorization, each item $j$ is mapped to a vector $v_j \in R^k$ ($k$ is the number of latent factors,) and each user $i$ is mapped to a vector $u_i \in R^k$. Each element $v_j^{k'}$ represents the feature $k'$ in the product $j$, and each element $u_i^{k'}$ represents the interest of the user $i$ in that feature. Inner product $u_i^T v_j$ estimates ratings of user $i$ to product $j$, and the equation of MF predictor is represented by,

$$\hat{r}_{i,j} = \mu + u_i^T v_j, \tag{1}$$

where $\mu$ is average rating in the user-item matrix. In order to learn the model parameters, the following least square problem need to be solved,

$$\min_{v_i, u_j} \sum_{(i,j) \in \Omega} \left( r_{i,j} - \mu - u_i^T v_j \right)^2 + \lambda \left( \|u_i\|^2 + \|v_j\|^2 \right). \tag{2}$$

In (2), $\Omega$ is the set of user-item pairs, which their rating is registered in the system and $\lambda$ is the regularization parameter, which is usually estimated through cross-validation. In order to find optimal solution of the above equation, Stochastic Gradient Descent (SGD) or Alternative Least Square (ALS) are used [19]. In recent years, different models have been proposed to improve the SVD model, among all SVD++[18], Bayesian probabilistic matrix factorization (BPMF)[20] and parametric probabilistic matrix factorization (PPMF)[21] can be mentioned.

One approach to improve model-based methods is to use user's reviews [22]-[24]. In [25] the authors considered demographic-based personal characteristics and features extracted from reviews to present a customized version of MF called matrix factorization with user attributes (MFUA). McAuley et al. [22] proposed hidden factors and hidden topics (HFT) model, in which the user's reviews are used to model the relationship between non-explicit interests of the user and non-explicit features of the products. HFT combined hidden factors learned from the user-product matrix with hidden topics learnt from reviews and has offered more accurate recommendations compared to the traditional MF model. For more information about recommender systems which employ user reviews Chen et al. [26] is a valuable survey.

Recent approach to deal with sparsity is by using cross-domain recommendation. Berkovsky et al. [27] have defined cross-domain recommendation as a technique that employs auxiliary data such as user's ratings, reviews contents [22], [28], contextual information [29], [30], social or information networks [31], [32] and additional feedbacks [33], [34] in one or more domains (auxiliary domains) to improve the recommendation accuracy in the target domain. Cantador et al. [7] considered four scenarios for cross-domain recommender system, based on data overlap between auxiliary and target domains including full overlap, user overlap, item overlap and no overlap. Approaches that do not consider overlapping between users, usually try to find a cluster-level rating pattern among different domains. Li et al. [1] proposed the cross-domain collaborative filtering method via codebook transfer (CBT) and increased the recommendation accuracy by building a codebook from dense auxiliary rating matrix and transferring it to sparse target ratings matrix. Sheng Gao et al. [35] extend the CBT model by relaxing hard membership constraint on user/item groups and considering domain's specific codebooks. Among approaches which considered overlapping between auxiliary and target domains, user overlapping approaches are more common. Hwangboa and Kim et al. [36] investigated different recommendation algorithms on real world datasets from Korean fashion companies. They show that ratio of overlapping between users and items of auxiliary and target domain has significant impact on recommendation accuracy, and high overlap domains has better accuracy in recommendations.

Zhang et al. [37] introduced cross-domain recommender system with consistent information transfer (CIT). They used a tri-factorization method along the domain adaptation techniques to generate consistent user/item latent groups in the source and target domains, and investigate what knowledge to transfer and how to effectively transfer that knowledge from the source domain to the target domain. Chen et al. Proposed TLRec[38], a transfer learning algorithm based on empirical prediction error, smoothness and regularization of user and item latent vectors for cross-domain recommendations on partially overlapped users. Since in most available datasets, correspondence between users in different domains is not specified, they considered movie genres as different domains and split the MovieLens dataset[39] into 4 domains including crime, comedy, drama and action, and selected a 1000 users and a 1000 items randomly for the experiment. Their experiments show effectiveness of the TLRec method.

Collective matrix factorization(CMF) is a powerful approach to simultaneously factorize several related matrices, which could be used in cross-domain recommender systems. The main CMF algorithm proposed in [40] jointly factorized the user-item matrix along with the user attributes and item attributes matrices. [41] proposed a convex formulation for CMF and [42] proposed an enhanced version of collective matrix factorization by considering different weight and regularization hyperparameters for each factorization section and independent parts for factorization matrices. Their empirical study on MovieLens dataset shows significant improvement over the traditional CMF.

Sahebi et al. [8], proposed a cross-domain recommender system algorithm based on Canonical Correlation Analysis (CCA).

The CCA has been previously used to obtain the relationship between different data such as text and image in single domain problem. For instance, in [2], CCA has been used to find a relationship between music tracks and human movements. Also in [43] CCA has been used to estimate ratings based on review texts and their semantic analysis. However, [8] is the first study which has used CCA for cross-domain recommendation. In [8], basic vectors $a$ and $b$ which maximize canonical correlation between rows of source and target user-item matrices are computed. Using these vectors, it can be estimated how item's ratings in the source domain affects the rating of items in the target domain. Sahebi et al. [8] proposed to predict user-item matrix in the target domain using the following equation:

$$\hat{R}^{(t)} = a\rho b^T R^{(s)}. \tag{3}$$

In this equation, $a$ and $b$ are canonical correlation vectors and $\rho$ is the corresponding canonical correlation. Their experiments on Yelp dataset has shown that the proposed method can improve cross-domain recommendation accuracy.

In recent years several deep learning based recommender system had been presented. Most of them focused on extracting content-based features from item's related contents such as item's text, descriptions or images [44]-[47]. For instance [44] proposed a deep text recommendation method that provides vector representation of the items content using deep recurrent neural networks by encoding the text sequent into a latent vector using gated recurrent units trained end-to-end on collaborative filtering task for scientific paper recommendation. [45] proposed a dual-regularized matrix factorization with deep neural networks to effectively exploiting descriptions of both users and items in matrix factorization. Their experiment shows improvement in the accuracy of rating prediction on Amazon and Yelp datasets. As most deep learning recommender systems are content-based, they fall beyond the scope of this paper and we will not focus on them. For more information on deep learning based recommender systems the readers could refer to [48].

## 3    THE PROPOSED METHOD

A recommender system is an information filtering system which tries to predict users rating for an item by collecting user's preference information. In this paper, user and product sets of a recommender system are shown with $\mathbb{U}$ and $\mathbb{V}$, respectively, and it is assumed that users' ratings are stored in a sparse user-item matrix $R$ in which element $r_{i,j}$ represents rating of user $i$ to product $j$, as shown in Fig. 1. Set of user-item pairs which have registered ratings in the recommender system and have a value in matrix $R$ are represented by $\Omega$. The set of products rated by user $i$ are represented by $\mathbb{V}_i$ and set of users which have rated products $j$ are represented by $\mathbb{U}_j$. By knowing $R$, $\mathbb{U}$ and $\mathbb{V}$, the objective of the recommender system becomes to estimating matrix $\hat{R}$ such that is closest to $R$ and has no missing values.

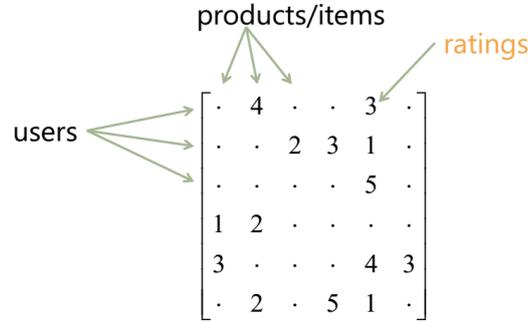

Fig. 1.  User-item matrix in a recommender system.

The purpose of cross-domain recommender system is to offer items in the target domain to all users in the source domain. To represent variable and matrices, belonging to each domain, we use superscript in parenthesis. For example, $R^{(s)}$ and $R^{(t)}$ represent user-item matrices in source and target domains respectively. In a cross-domain recommender system, with knowing user-item matrices $R^{(s)}$ and $R^{(t)}$ the purpose is to predict missing ratings in $R^{(t)}$ from users in $U^{(s)}$. As mentioned in the previous section, the proposed method is based on generalized canonical correlation analysis (GCCA). To present the proposed we briefly explain CCA and GCCA.

### 3.1    Canonical Correlation Analysis

The purpose of canonical-correlation analysis is to find linear transformation of two multi-dimensional $x$ and $y$ which have maximum correlation with each other [49]. In CCA, basis vectors $a_1 \in R^{k_x}$ and $b_1 \in R^{k_y}$ are estimated for two multi-dimensional variables $x \in R^{k_x}$ and $y \in R^{k_y}$ such that correlation between projection of variables on the basis vectors are mutually maximized and correlation matrix of variables is diagonal. Assume $a_1$ and $b_1$ are basis vectors which maximize correlation between canonical variables $x' = a_1^T x$ and $y' = b_1^T y$. In other words, we have:

$$\max_{a_1 \in R^{k_x}, b_1 \in R^{k_x}} cor(x', y') = \max_{a_1 \in R^{k_x}, b_1 \in R^{k_x}} \frac{E[a_1^T xy^T b_1]}{\sqrt{E[a_1^T xx^T a_1] E[b_1^T yy^T b_1]}} = \max_{a_1 \in R^{k_x}, b_1 \in R^{k_x}} \frac{a_1^T C_{xy} b_1}{\sqrt{a_1^T C_{xx} a_1 b_1^T C_{yy} b_1}} \tag{4}$$

In this equation, $C_{xx}$, $C_{yy}$ are correlation matrices of $x$ and $y$ and $C_{xy}$ is mutual correlation of $x$ and $y$. Vectors $a_1$ and $b_1$ which maximize the above equation, are the primary basic vectors of canonical correlation. It should be mentioned that for any of the next canonical variable $x'_i = a_i^T x$ and $y'_i = b_i^T y$, it should be uncorrelated with previous variables $x'_j = a_j^T x$ and $y'_j = b_j^T y$ ($j < i$). In other words:

$$\begin{cases} E[x'_i x'^T_j] = a_i^T C_{xx} a_j = 0 \\ E[y'_i y'^T_j] = b_i^T C_{yy} b_j = 0 \quad ; \; for\ i \neq j \\ E[x'_i y'^T_j] = a_i^T C_{xy} b_j = 0 \end{cases} \tag{5}$$

It can be shown the next pairs of canonical basis $\{(a_i, b_i) | i < \min(k_x, k_y)\}$ can be found by solving the following eigenvalue equation [49]:

$$\begin{cases} C_{xx}^{-1} C_{xy} C_{yy}^{-1} C_{yx} a_i = \rho^2 a_i \\ C_{yy}^{-1} C_{yx} C_{xx}^{-1} C_{xy} b_i = \rho^2 b_i \end{cases}, \tag{6}$$

where $\rho^2$ is the squared canonical correlation and eigenvectors $a_i$ and $b_i$ are normalized canonical correlation basis vectors.

### 3.2 Generalized canonical correlation analysis

Generalized canonical correlation analysis (GCCA) is the extended version of CCA that could be applied on more than two spaces. Unlike CCA which has a simple algebraic solution, GCCA is a difficult optimization problem, which does not have a simple solution. Several generalized canonical correlation analysis have been proposed, some of them are discussed and compared in [50], among them the MAX-VAR GCCA proposed in [51] is well-known. MAX-VAR GCCA tries to find a linear map which creates identical latent representations from different views instead of finding a linear map which creates maximum correlation. MAX-VAR GCCA is straightforward and its solution is based on an eigen equation.

Assuming that $\{X^{(d)}; d \in \mathcal{D}\}$ are a set of mean centered matrices of different domains (views), and each row $x_{i,:}^{(d)}$ of matrix $X^{(d)}$ represents features of sample $i$ in domain $d$, the objective function of MAX-VAR GCCA defines as:

$$\underset{G, W^{(d)}}{\operatorname{argmin}} \sum_{d \in D} \left\| G - X^{(d)} W^{(d)} \right\|_F^2 \quad (7)$$

$subject\ to\ G^T G = I,$

where $W^{(d)} \in R_{k_d \times k}$ correspond to the $k$ canonical component of view $d$, and $G$ is a $L \times K$ group configuration matrix. Since in several situations we might have missing samples in some domains, Van De Velden and Bijmolt [52], modified the objective function of MAX-VAR GCCA to support missing samples:

$$\underset{G, W^{(d)}}{\operatorname{argmin}} \sum_{d \in D} \left\| K^{(d)} \left( G - X^{(d)} W^{(d)} \right) \right\|_F^2 \tag{8}$$

$subject\ to\ G^T K G = nI,$

where $n = |D|$ is the number of domains and $K^{(d)}$ is a diagonal selection matrix in which $K_{i,i}^{(d)} = 1$ if sample $i$ is observed in domain $d$ and $K_{i,i}^{(d)} = 0$ if it does not. $K = \sum_{d=1}^{n} K^{(d)}$ is the sum of selection matrices. Assuming that matrices $X^{(d)^T} X^{(d)}$ are nonsingular, $W^{(d)}$ can be calculated as:

$$W^{(d)} = X^{(d)\dagger} G = \left( X^{(d)^T} X^{(d)} \right)^{-1} X^{(d)^T} G. \tag{9}$$

Let us define:

$$P^{(d)} = X^{(d)} \left( X^{(d)^T} X^{(d)} \right)^{-1} X^{(d)^T}, \tag{10}$$

then matrix $M$ is:

$$M = \sum_{d \in D} P^{(d)}. \tag{11}$$

The group configuration $G$ is determined by solving the spectral decomposition of sum of projection matrices of different domains:

$$MG^* = G^* \Lambda, \tag{12}$$

where $G^*$ is an orthonormal matrix and $\Lambda$ is a diagonal matrix. Main diagonal elements of $\Lambda$ are the largest eigenvalues of $M$ and $G^*$ is a $L \times k$ matrix of corresponding orthonormal eigenvectors. Group configuration $G$ can be calculated as $G = \sqrt{n} K^{-\frac{1}{2}} G^*$.

### 3.3 GCCA with inverse sum of selection matrices

The constraint defined by $G^T K G = nI$ in (8) forces samples which are relatively infrequent to observed to have larger weight compared to the samples which are relatively often observed. This helps infrequently observed samples to affect more on the learning hidden space features. This property might be helping in GCCA, but in transferring knowledge between domains, samples which have been observed in most domains have more cross-domain information and can specify the relationship between features in different domains better than less observed samples, therefore they should have higher impact. In order to solve this problem, we proposed to inverse the sum of selection matrices in constraint of (8) to increase the effect of samples which are relatively often observed. This new approach will increase the effect of relatively often observed sample on the objective function by letting their corresponding row in $G$ to have larger values. We call the proposed method generalized canonical correlation by inverse sum of selection matrices (GCCA-ISSM) and define its objective function as:

$$\underset{G,W^{(d)}}{\mathrm{argmin}} \sum_{d\in D}\left\|K^{(d)}\left(G - X^{(d)}W^{(d)}\right)\right\|_F^2 \tag{13}$$

$$\text{subject to } G^T K^{-1} G = nI.$$

Similar to GCCA, it can be shown that by defining $P^{(d)} = k^{(d)} - X^{(d)}\left(X^{(d)^T}X^{(d)}\right)^{-1}X^{(d)^T}$, the solution to the above equation is obtained using the same eigenequation as (12). Also matrix $G$ can be found as $G = \sqrt{n}K^{\frac{1}{2}}G^*$.

Memory requirement for caching matrix $M$ brings serious limitation in learning GCCA and GCCA-ISSM. We use technique in [53] to propose a fast and memory efficient version of GCCA-ISSM called Fast-GCCA-ISSM. Let consider the rank-m SVD of $X^{(d)} = A^{(d)}S^{(d)}B^{(d)^T}$, where $S^{(d)}$ is a diagonal matrix with $m$ largest singular values of $X^{(d)}$ and $A^{(d)}$, $B^{(d)}$ are the corresponding left and right singular vectors. then, $\tilde{P}^{(d)}$ is defined using rank-$m$ SVD of $X^{(d)}$ as follows:

$$\tilde{P}^{(d)} = A^{(d)}S^{(d)}\left(r^{(d)}I + S^{(d)}S^{(d)^T}\right)^{-1}S^{(d)}A^{(d)^T} = A^{(d)}T^{(d)}T^{(d)^T}A^{(d)^T}. \tag{14}$$

In which $r^{(d)}$ is a small constant noise and $T^{(d)} \in R^{n\times n}$ is a diagonal matrix such that $T^{(d)}T^{(d)^T} = S^{(d)}(r^{(d)}I + S^{(d)}S^{(d)^T})^{-1}S^{(d)}$. Assuming $\tilde{M} = [A^{(d_1)}T^{(d_1)} \dots A^{(d_n)}T^{(d_n)}]$, sum of projection matrices is written as $M = \tilde{M}\tilde{M}^T$.

The group configuration matrix $G$ are obtained from left singular vectors of $K^{\frac{1}{2}}\tilde{M}$. The memory requirement of Fast-GCC-ISSM is in the order of $O(Lm)$ compare to $O(L^2)$ in GCCA-ISSM.

ALGORITHM 1
GCCA-ISSM CROSS-DOMAIN RECOMMENDER SYSTEM

---

1: // $L$: Number of users.
2: // $R^{(d)}$: ratings matrix for domain $d$.
3: // $K^{(d)}$: selection matrix for domain $d$
4: For each domain $d$ in $D$:
5:     Initialize $U^{(d)}$ and $V^{(d)}$ with normal distribution.
6: Do until convergence off $err$:
7:     $M \leftarrow \bar{0}$
8:     For each domain $d$ in $D$:
9:         Update $U^{(d)}$ and $V^{(d)}$ using SGD matrix factorization on $R^{(d)}$
10:         $P^{(d)} \leftarrow K^{(d)} - U^{(d)}\left(U^{(d)^T}U^{(d)}\right)^{-1}U^{(d)^T}$
11:         $M \leftarrow M + P^{(d)}$
12:     $G^*, \Lambda \leftarrow eig(M)$
13:     $G \leftarrow \sqrt{n}K^{-\frac{1}{2}}G^*$
14:     For each domain $d$ in $D$:
15:         $W^{(d)} \leftarrow \left(X^{(d)^T}X^{(d)}\right)^{-1}X^{(d)^T}$
16:         $U^{(d)} \leftarrow GW^{(d)^{-1}}$
17:     $\hat{R}^{(t)} \leftarrow GW^{(t)^{-1}}V^{(t)}$
18:     $err \leftarrow \left\|R^{(t)} - \hat{R}^{(t)}\right\|_2$
19: return $\hat{R}^{(t)}$

---

### 3.4    GCCA-ISSM cross-domain Recommender System

The cross-domain recommender system proposed in this paper is an iterative procedure that alternates between two steps: Stochastic gradient descent matrix factorization on user-items matrix and generalized canonical correlation analysis on user's latent factors obtained from matrix factorization. The first step tries to reduce the ratings prediction error in each domain using matrix factorization while the second steps tries to extract domain independents features from user's latent factors matrices using canonical correlation analysis and reconstructing the user's latent factors in each domain to improve the ratings prediction of users, specially the cold-start users. In the first step the SGD matrix factorization is applied on user-items matrix of each domain $(d)$ from auxiliaries and target domains to estimate the user's and item's latent factors $(U^{(d)},V^{(d)})$. In the second step the canonical correlation algorithm (GCCA or GCCA-ISSM) is applied on user's latent factors $(U^{(d)})$ of all domains and the canonical component $W^{(d)}$ and domain independent user's latent features $G$ is calculated. Then the user's latent factors of each domain $(U^{(d)})$ is re-estimated by $U^{(d)} = GW^{(d)^{-1}}$ which inferred from MAX-VAR GCCA objective function in equation (7). Finally the user-item matrix of target domain is predicted using the estimated user's latent factors from step two and the mean square error of the predicted matrix is calculated. The algorithms iterate between these steps until the converges and while the mean square error on the validation set is decreasing. Algorithm 1, presents the pseudocode of the proposed method for cross-domain recommendation. In most of our experiments the proposed algorithm converged in less than 10 iterations.

## 4 EXPERIMENTS WITH SIMULATED DATA AND DISCUSSIONS

In this section, we investigate properties of GCCA and GCCA-ISSM by conducting a simulation experiment. We use the Algorithm 2 which is adopted from [52] to generate synthetic and investigate the accuracy of GCCA and GCCA-ISSM in reconstructing data matrices.

ALGORITHM 2
GENERATING SYNTHETIC DATA

| |
|---|
| 1:    // L: number of sample(Users) |
| 2:    // M: latent space dimension |
| 3:    // s: sparsity |
| 4:    // $\sigma = 0.1$ : noise |
| 5:    Create orthonormal random $G_{L \times M}$ |
| 6:    For each domain: |
| 7:      Choose $A_{M \times M^{(d)}}^{(d)}$ from $N(0,1)$ |
| 8:      Choose $B_{L \times M}^{(d)}$ from $N(0,1)$ |
| 9:      $X_{complete}^{(d)} \leftarrow G.A^{(d)} + \sigma \times B$ // complete data matrices |
| 10:     $X_{sparse}^{(d)} \leftarrow X^{(d)}$ // sparse data matrices |
| 11:     $K^{(d)} = I$ (Identity Matrix) |
| 12:     missing_rows = choose $s \times L$ random number from $N(0,1)$ |
| 13:     for $i$ in missing_rows |
| 14:       $K_{i,i}^{(d)} \leftarrow 0$ |
| 15:       $X_{sparse\ i,:}^{(d)} \leftarrow \vec{0}$ //(replace row $i$ of $X_{sparse}^{(d)}$ by zero) |
| 16:    return $K^{(1..n)}, X_{sparse}^{(1..n)}$ |

In this experiment, influence og three parameters including number of domains ($n$), data sparsity ($s$) and number of rows of the observation matrices ($L$) are investigated. For each algorithm we used Mean Square Error (MSE) of reconstructed matrices for evaluation. For each parameter setting the experiment is repeated 1000 times and the results are obtained by averaging the MSE of all experiments. As in different experiments, absolute value of MSE for GCCA and GCCA-ISSM are different and depend on initialization parameters in generating synthetic data, the improvement ratio of MSE is a better measure to evaluate the results.

Fig. 2 shows the effect of data matrices sparsity on MSE and improvement ratio of GCCA and GCCA-ISSM. This experiment performed in two scenarios with 4 and 8 number of domains. In Fig. 2, blue and red lines belong to GCCA and GCCA-ISSM respectively. Dashed lines show experiment MSE of reconstructing data matrices, and solid lines show MSE of reconstructing missing rows. Green line in the bottom diagrams show improvement ratio of GCCA-ISSM against traditional GCCA.

As can be seen in the results, standard GCCA performs better than the proposed GCCA-ISSM method in reconstructing data matrices, but GCCA-ISSM obtained better results in reconstructing missing rows and improved the MSE from 20 to 50 percent. The outcome is in line with our objective in cross-domain recommender system, and shows that the proposed method's ability to reconstruct missing rows in observation matrices is better than traditional GCCA. In addition, this experiment shows by increasing sparsity of rows, MSE of reconstructing missing rows increase for both GCCA and GCCA-ISSM, but GCCA-ISSM performs a lot better than GCCA, and its improvement ratio is almost stable.

To examine the scalability of the proposed algorithm in the second experiment we investigate effect of number of matrix rows ($L$) on reconstruction accuracy of GCCA and GCCA-ISSM on missing rows. In this experiment, it is assumed we have 2 domains and 40 percent of the matrix rows for each domain are missing (sparsity=0.4). Fig. 3 presents the experiment results. In this figure, the black solid line shows default MSE of simulation by considering the noise, which is the best result that could be obtained. Fig. 3 shows the reconstruction ability of GCCA-ISSM on missing rows is always better than GCCA, and as the number of matrix rows increase, accuracy of GCCA-ISSM improves more than GCCA. Interesting point in this experiment is that by increasing the number of matrices rows to more than 2500, MSE of GCCA-ISSM in reconstructing complete data matrices would become better than GCCA. Obtained result shows, even in large datasets the ability of GCCA-ISSM in reconstructing missing rows is much better than traditional GCCA.

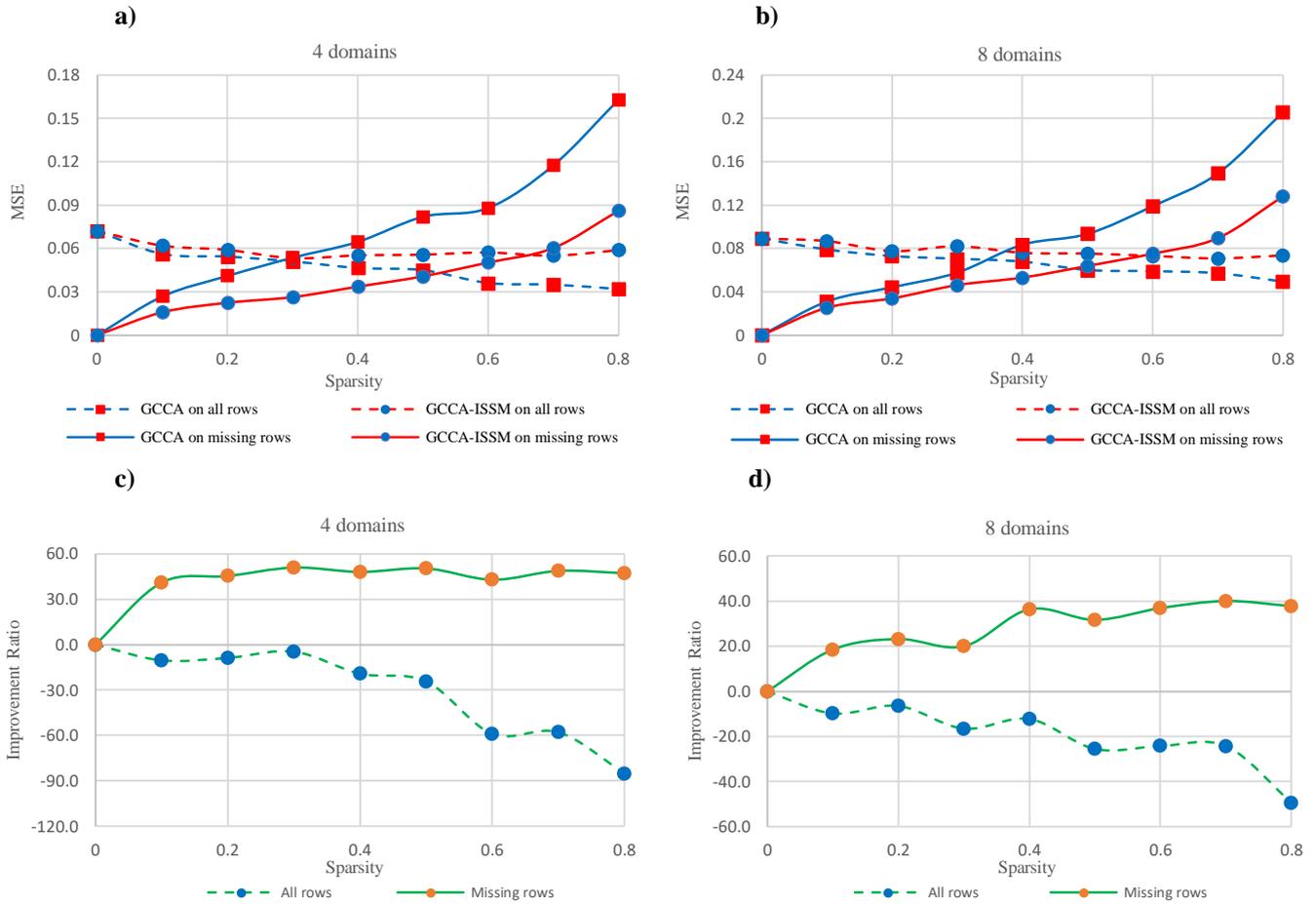

Fig. 2. Comparison of GCCA and GCCA-ISSM on synthetic data with different sparsity.
(a),(b) Reconstruction MSE of GCCA and GCCA-ISSM for 4 and 8 domains.
(c), (d) improvement ratio of GCCA-ISSM compared to GCCA for 4 and 8 domains.

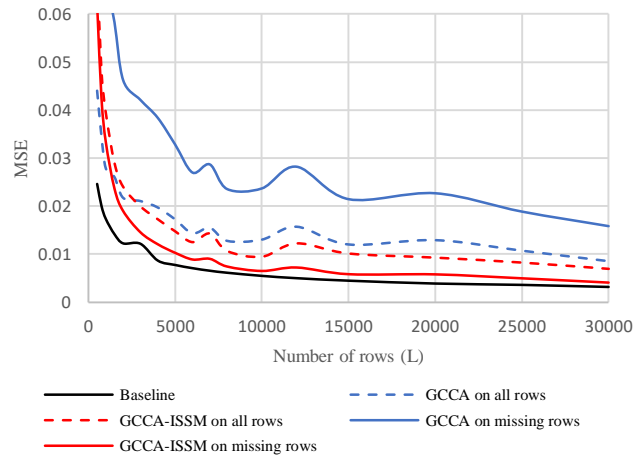

Fig. 3. Performance of GCCA and GCCA-ISSM with growth of number of samples.

## 5 EXPERIMENT WITH RECOMMENDER SYSTEM DATA AND DISCUSSIONS

In this section, we describe our experimental setup and the evaluation results of the proposed method on real datasets.

### 5.1 Datasets

In recent years, several public datasets have been used for evaluating recommender systems, among them Amazon[54], Netflix[55], Yelp[56], MovieLens[39], [38] and TripAdvisor[57] are the most well-known ones. Since Amazon dataset includes information from items in different domains, it is suitable for evaluating cross-domain recommendation algorithm. However, other datasets have also been used. For example, in [38] by simulating domains using movie genres, MovieLens has been used for evaluation. In this paper, we used the most recent version of Amazon dataset introduced in [54] and the 1 million ratings Movielens dataset from [39].

The Amazon dataset includes reviews and metadata of the products offered in online Amazon market from 1996 to 2014. Reviews include information such as username, user's rating to the items, review texts, summaries, usefulness and etc. Many of the proposed cross-domain methods have tested their results only on book and movie domains[35], [38], [58], [59]. In this paper, we increase the number of domains in experiments to 6 domains including "Books", "Kindle,", " Movies", "CDs", "Digital Music" and "Video Games". These domains are selected based on the number of common users between domains and the accuracy of MF algorithms in single domain experiment. Due to the high sparsity of Amazon dataset and in order to increase reliability of our experiments, only users who have at least 5 ratings in each domain are selected.

MovieLens dataset contain 1 million anonymous ratings of approximately 3,900 movies in different genres made by 6,040 users who joined MovieLens in 2000. The dataset ensures that each users has at least 20 ratings. We performed our experiments on the top four genre including "Action", "Comedy", "Drama" and "Thriller" based on the number of ratings in each genre. Table 2 presents more details about the datasets.

Experiments are conducted using 5-fold cross validation; 3 folds for training, 1 fold for validation and 1 fold for test. Depending on the experiment, we removed all or some ratings of test's users from training data to simulate the cold-start. Hence, learning latent factors of the test users are only feasible through transferring knowledge from auxiliary domains.

<div align="center">TABLE 2<br>DATASET STATISTICAL INFORMATION</div>

| Amazon[54] | | | | | Movelens[39] | | | | |
|---|---|---|---|---|---|---|---|---|---|
| **Domain** | **#users** | **#items** | **#ratings** | **sparsity** | **Domain** | **#users** | **#items** | **#ratings** | **sparsity** |
| Books | 603,668 | 367,982 | 8,898,041 | 0.9999 | Action | 6,012 | 495 | 257,457 | 0.9134 |
| Kindle | 68,223 | 61,934 | 982,619 | 0.9997 | Comedy | 6,031 | 1,163 | 356,580 | 0.9491 |
| Movies | 123,960 | 50,052 | 1,697,533 | 0.9997 | Drama | 6,037 | 1,493 | 354,529 | 0.9606 |
| CD | 75,258 | 64,443 | 1,097,592 | 0.9997 | Thriller | 5,989 | 485 | 189,680 | 0.9346 |
| Videogames | 24,303 | 10,672 | 231,780 | 0.9991 | | | | | |
| Digital Music | 5,541 | 3,568 | 64,706 | 0.9967 | | | | | |

## 5.2 Baselines

We compare the following methods and baselines to evaluate the accuracy and performance of the proposed methods.

a)  Offset: The offset is the average across all ratings in training data. It is the best possible constant predictor.

b)  MF: This is the standard single domain matrix factorization[17] using stochastic gradient descent optimization.

c)  CBT: The cross-domain collaborative filtering method via codebook transfer(CBT) presented in [1].

d)  CMF: The most recent version of collaborative matrix factorization presented in [42] which has better prediction accuracy in cold-start situations.

e)  MF-GCCA: the proposed iterative cross-domain algorithm using generalized canonical correlation analysis.

f)  MF-GCCA-ISSM: The proposed iterative cross-domain algorithm using generalized canonical correlation analysis with inverse sum of selection matrix. The GCCA-ISSM is expected to perform better than GCCA in cold-start situations.

g)  MF-Fast-GCCA-ISSM: The proposed cross-domain iterative method using fast and memory efficient Fast-GCCA-ISSM algorithm.

h)  (PCA-LDA)-GCCA-ISSM: Based on the experiment setup different combination of auxiliary data may be used for cross-domain prediction.

## 5.3 Cross-domain experiment

The purpose of this experiment is to investigate the accuracy of proposed MF-GCCA and MF-GCCA-ISSM algorithms in cross-domain rating prediction for cold-start users and compare the results with the baseline methods. To simulate the cold-start problems, first we define "c-cold-start users" as users from auxiliary domains that had at most $c$ ratings in the target domain. The experiment had done in two parts. In the first part to compare the accuracy of proposed MF-GCCA and MF-GCCA-ISSM on solving cold-start users rating prediction, for each pair of selected domains in the Amazon and MovieLens datasets, one domain was selected as auxiliary and the other as target domain. We considered 50 latent factors for learning MF($k^{(d)} = 50$) and 75 latent factors for common space in MF-GCCA and MF-GCCA-ISSM($k = 75$). To evaluate the proposed method with the baselines the mean square error of ratings prediction for 0-cold-start users had been used. We compare the results with Offset (since there is no ratings for the test users in the training data of target domain, the Offset algorithm would produce the same results as MF), CBT and CMF. Table 3 and Table 4 presents the MSE and improvement ratio of each methods on Amazon and MovieLens datasets. On Amazon dataset for each target domain we present two auxiliary domains that create higher improvement.

As can be seen in Table 3 and 4, both MF-GCCA and MF-GCCA-ISSM improve cross-domain recommendation accuracy. While MF-GCCA could not reduce the prediction accuracy in few domains like Movies and Videogames in Amazon dataset, the MF-GCCA-ISSM outperform the baseline methods in all domains of both datasets. The experiment on Amazon dataset shows that the pair of (Kindle, Book) and (DigitalMusic, CD) has highest improvement ratio that means they have more related features compared to other domains. "Videogames" has a lower improvement ratio among all.

Since Movielens is denser than Amazon (the auxiliary domain is denser) the MF-GCCA-ISSM algorithms results better improvement. On average it improve over 30% against Offset method and more than 20% against CBT and CMF. Both CBT and CMF depends on existence of user ratings in target domain. If test users had no ratings in the target domain, CBT and CMF performance would be limited.

Furthermore, in second part we investigated a fair comparison between MF, CMF, CBT and our proposed approach on $\{0\ to\ 25\}$-cold-start users ratings. Based on the predicted accuracy we selected the (Target=Action, Auxiliary=Drama) pair from Movielens and (Target=Books, Auxiliary=Movies) pair from Amazon. For $c \in \{0\ to\ 25\}$ we let the test users to have at most $c$ ratings in the target domain. For each $c$, the experiment has executed independently and the results are presented in Fig 4 . It could be seen that almost for any value of $c$ the MF-GCCA-ISSM algorithm has the lowest MSE and outperformed the baselines. For lower $c$ the improvement of MF-GCCA-ISSM against the single domain MF is significant. As $c$ increased the user's cold-start problem reduced hence the improvement of the MF-GCCA-ISSM and others cross-domain algorithm against the single domain MF decreased. In both datasets for $c < 6$ CMF has lower prediction error against single domain MF. While for $c \geq 6$ the single domain MF has exceed the CMF. This shows that CMF improvement in cross-domain prediction for cold-start users is limited to users that had very few ratings in target domain. In both datasets the cross-domain prediction MSE of CBT was worse than the single domain MF. The reasons are: (1) The CBT algorithm does not use the relationship of common users in auxiliary and target domain. (2) The CBT algorithm needs the auxiliary domain to be denser than the target domain to perform better than single domain MF. But in real cross-domain scenario like our experiments the density of auxiliary and target domain is the same.

TABLE 3
CROSS-DOMAIN RESULTS OF PROPOSED METHOD FOR COLD-START USERS ON AMAZON DATASET

| Domains | | Mean Square Error | | | | | Improvement over Offset (%) | | | |
|---|---|---|---|---|---|---|---|---|---|---|
| Target Domain | Auxiliary Domain | Offset (MF) | CBT | CMF | MF-GCCA | MF-GCCA-ISSM | CBT | CMF | MF-GCCA | MF-GCCA-ISSM |
| CD | Digital Music | 1.006 | 1.0056 | 0.9933 | 0.9477 | **0.9021** | 0.06% | 1.28% | 5.80% | **10.34%** |
| CD | Movies | 1.0440 | 1.0524 | 1.0414 | 1.0242 | **0.9984** | -0.80% | 0.25% | 1.90% | **4.37%** |
| Movies | Books | 1.3663 | 1.3428 | 1.3651 | 1.4206 | **1.3169** | 1.73% | 0.09% | -3.97% | **5.91%** |
| Movies | CD | 1.2652 | 1.2664 | 1.2652 | 1.3215 | **1.1824** | -0.09% | 0.36% | -4.45% | **6.54%** |
| Kindle | Books | 0.8203 | 0.8311 | 0.8177 | 0.7518 | **0.7474** | -1.32% | 0.32% | 8.34% | **8.89%** |
| Kindle | Movies | 1.0017 | 1.0114 | 1.001 | 1.001 | 1.001 | -0.96% | 0.10% | 0.10% | 0.10% |
| Books | Kindle | 0.7693 | 0.7587 | 0.7455 | 0.7398 | **0.7272** | 1.39% | 3.09% | 3.83% | **5.48%** |
| Books | Movies | 1.1263 | 1.1225 | 1.1254 | 1.1238 | **1.1221** | 0.35% | 0.08% | 0.22% | **0.38%** |
| Digital Music | CD | 1.1405 | 1.1763 | 1.1059 | 0.9946 | **0.9519** | -3.14% | 3.03% | 12.79% | **16.53%** |
| Digital Music | Movies | 1.1807 | 1.1772 | 1.1576 | 1.1595 | **1.1444** | 0.30% | 1.96% | 1.80% | **3.08%** |
| Videogames | Digital Music | 1.2933 | 1.3302 | 1.3228 | 1.3331 | 1.4211 | 0.14% | 0.70% | -0.08% | 0.17% |
| Videogames | Movies | 1.4313 | 1.4552 | 1.4271 | 1.4370 | 1.3297 | -1.67% | 0.29% | -0.40% | 0.71% |

TABLE 4
CROSS-DOMAIN RESULTS OF PROPOSED METHOD FOR COLD-START USERS ON MOVIELENS DATASET

| Domains | | Mean Square Error | | | | | Improvement over Offset | | | |
|---|---|---|---|---|---|---|---|---|---|---|
| Target Domain | Auxiliary Domain | Offset (MF) | CBT | CMF | MF-GCCA | MF-GCCA-ISSM | CBT | CMF | MF-GCCA | MF-GCCA-ISSM |
| Action | Comedy | 1.3340 | 1.2834 | 1.2519 | 1.2176 | **0.8700** | 3.79% | 6.15% | 8.72% | **34.78%** |
| Action | Drama | 1.2398 | 1.3361 | 1.1575 | 1.1094 | **0.8044** | -7.77% | 6.64% | 10.52% | **35.12%** |
| Action | Thriller | 1.2949 | 1.2996 | 1.1778 | 1.1319 | **0.7480** | -0.43% | 8.98% | 12.53% | **42.19%** |
| Comedy | Action | 1.2660 | 1.1666 | 1.2088 | 1.1776 | **0.8819** | 7.85% | 4.52% | 6.98% | **30.34%** |
| Comedy | Drama | 1.2357 | 1.3231 | 1.1738 | 1.1153 | **0.8534** | -7.08% | 5.00% | 9.74% | **30.94%** |
| Comedy | Thriller | 1.2269 | 1.2420 | 1.1738 | 1.2829 | **0.8990** | -1.23% | 4.33% | -4.56% | **26.72%** |
| Drama | Action | 1.0710 | 1.0955 | 1.0190 | 0.9551 | **0.7924** | -2.29% | 4.85% | 10.82% | **26.01%** |
| Drama | Comedy | 1.1125 | 1.0938 | 1.0614 | 0.9612 | **0.7910** | 1.68% | 4.59% | 13.60% | **28.90%** |
| Drama | Thriller | 1.0368 | 0.9351 | 0.9894 | 1.0145 | **0.7976** | 9.80% | 4.57% | 2.14% | **23.07%** |
| Thriller | Action | 1.2275 | 1.2006 | 1.1292 | 0.9510 | **0.7416** | 2.19% | 8.01% | 22.52% | **39.58%** |
| Thriller | Comedy | 1.2420 | 1.2187 | 1.1712 | 1.1613 | **0.8334** | 1.88% | 5.70% | 6.50% | **32.89%** |
| Thriller | Drama | 1.2577 | 1.3359 | 1.1816 | 1.006 | **0.7881** | -6.22% | 6.05% | 19.94% | **37.34%** |

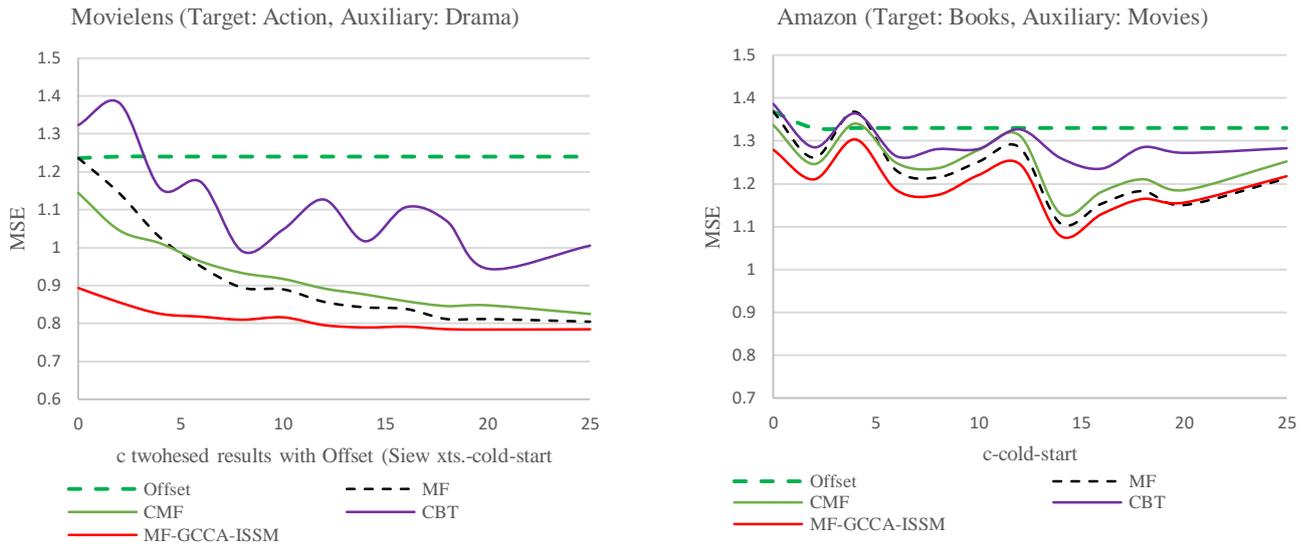

Fig. 4. Comparison of baselines and proposed method for different number of users in training data.

### 5.4    Cross-domain with multiple auxiliary domain

In this experiment we study the ability of proposed MF-GCCA-ISSM algorithm on using multiple auxiliary domains to improve the rating prediction accuracy of cross-domain recommendation. We perform the experiment on both Amazon and MovieLens dataset. From amazon dataset the "Movies", "Digital Music" and "Videogames" domains are selected as auxiliary and the "CD" domain is selected as target domain. From MovieLens dataset the "Comedy", "Drama" and "Thriller" genres are selected as auxiliary domain and the "Action" genre is selected as target domain. We performed the experiment using Offset, single domain MF and MF-GCCA-ISSM for different combination of auxiliary domain  Fig. 5 presents the MSE of rating predictions for 5-cold-start users in target domain in bar charts. This figure shows in Amazon dataset, the "Digital Music" and the "Movies" domain respectively had the most effects on reducing the MSE. The "Videogames" could not improve the results compared to the single domain MF but its combination with "Digital Music" and "Movies" domains results the most improvement in ratings prediction. In MovieLens dataset each of the auxiliary domains independently reduced the ratings prediction's MSE in target domain to less than 0.9 and the combination of all three auxiliary domains reduced the MSE to less than 0.8. In both datasets the best results obtained when all domains have been used as auxiliary data. These results shows that the proposed algorithm could be extended to situations with multiple auxiliary domains without losing accuracy and efficiency. Also a small difference between the best results of two auxiliary domains and three auxiliary domains shows the limitations of improving the recommendation by adding more auxiliary domains. As number of related auxiliary domains increased adding a new auxiliary domain would have less effect on improving the cross-domain recommendations.

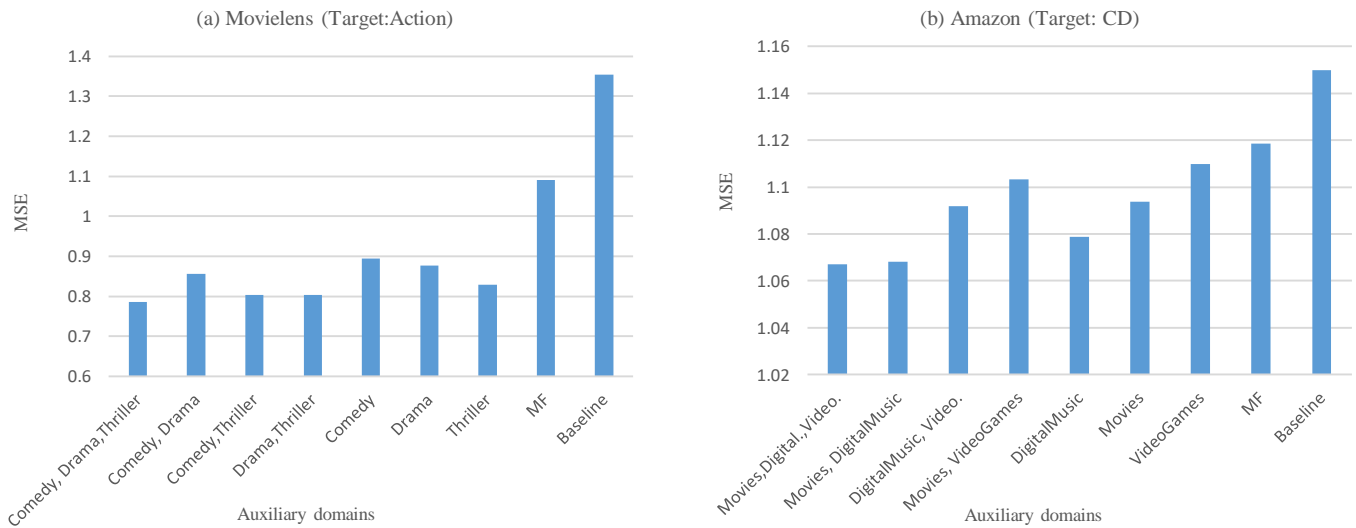

Fig. 5. Performance of MF-GCCA-ISSM on cross-domain recommendation with multi auxiliary domains.
(a) Action as target domain and Comedy, Drama and Thriller as auxiliary domains.
(b) CD as target domain and Movies, Digital Music and Videogames as auxiliary domains.

### 5.5 Using extra auxiliary data

The purpose of this experiment is to investigate performance of GCCA-ISSM in learning features from auxiliary data like demographic information, reviews and etc. In this experiment, we investigate three types of auxiliary data including MF, PCA and Latent Dirichlet Allocation(LDA). Also, different combinations of these auxiliary data are used in the experiments. For MF, we used the users latent factor learnt from standard matrix factorization like previous experiments. For PCA, instead of MF, we applied PCA on user-item matrix in the auxiliary domain and used principal components to learn latent factors of users in the target domain. For LDA, users' reviews text in auxiliary domain is considered as auxiliary data. We applied a topic modeling algorithm called Latent Dirichlet Allocation [60] to the user reviews in the auxiliary domain (We considered all reviews of a user as a document). Then we considered the distribution of topics used by users as latent factors of the target domain.

Table 5 presents the results of this experiment. It shows that all three auxiliary data (PCA, LDA and MF) could increase the cross-domain recommendation accuracy. Although in most cases when we use MF as auxiliary data, the improvement of prediction is more notable. This experiment shows that the proposed GCCA-ISSM algorithm is not limited to matrix factorization and it can transfer knowledge from different auxiliary data. Also, in most experiments the results shows that the combination of MF and LDA as auxiliary data gains the best results.

TABLE 5
CROSS-DOMAIN ACCURACY OF GCCA-ISSM COMBINATIONS WITH PCA AND LDA

| Target Domain | Auxiliary Domain | MSE/Imp. ratio | Baseline | PCA-GCCA-ISSM | LDA-GCCA-ISSM | MF-GCCA-ISSM | PCA-LDA-GCCA-ISSM | PCA-MF-GCCA-ISSM | LDA-MF-GCCA-ISSM | PCA-LDA-MF-GCCA-ISSM |
|---|---|---|---|---|---|---|---|---|---|---|
| CD | Digital Music | MSE | 1.0136 | 0.9847 | 0.9842 | 0.9136 | 0.9712 | 0.9259 | **0.9026** | 0.9087 |
| | | Imp. ratio | - | 2.85% | 2.90% | 9.87% | 4.18% | 8.65% | **10.95%** | 10.34% |
| CD | Movies | MSE | 1.0805 | 1.0604 | 1.0711 | 1.0428 | 1.0639 | **1.0359** | 1.045 | 1.0418 |
| | | Imp. ratio | - | 1.86% | 0.87% | 3.49% | 1.53% | **4.13%** | 3.25% | 3.59% |
| Movies | Digital Music | MSE | 1.3947 | 1.3748 | 1.3746 | **1.3435** | 1.3593 | 1.3521 | 1.3453 | 1.3516 |
| | | Imp. ratio | - | 1.43% | 1.44% | **3.67%** | 2.54% | 3.06% | 3.54% | 3.09% |
| Movies | CD | MSE | 1.3301 | 1.2951 | 1.2862 | 1.2752 | 1.2942 | 1.2836 | **1.2381** | 1.2505 |
| | | Imp. ratio | - | 2.63% | 3.30% | 4.12% | 2.70% | 3.49% | **6.91%** | 5.98% |
| Digital Music | CD | MSE | 1.1935 | 1.1543 | 1.1559 | 1.0365 | 1.1307 | 1.0079 | 1.0271 | **0.9946** |
| | | Imp. ratio | - | 3.28% | 3.15% | 13.15% | 5.26% | 15.55% | 13.94% | **16.66%** |
| Digital Music | Movies | MSE | 1.1193 | 1.0896 | 1.0887 | 1.0492 | 1.0724 | 1.0824 | **1.0427** | 1.0681 |
| | | Imp. ratio | - | 2.74% | 2.74% | 6.26% | 4.19% | 3.29% | **6.84%** | 4.57% |

TABLE 6
CROSS-DOMAIN ACCURACY AND MEMORY USAGE OF MF-Fast-GCCA-ISSM COMPARED TO MF-GCCA-ISSM

| min. # of ratings for each user | # of users | # of ratings | MF-GCCA-ISSM Memory (Estimated) | MF-Fast-GCCA-ISSM Memory (Estimated) | MF (MSE) | MF-GCCA-ISSM (MSE) | MF-Fast-GCCA-ISSM (MSE) | MF-GCCA-ISSM (Imp. Ratio) | MF-Fast-GCCA-ISSM (Imp. Ratio) |
|---|---|---|---|---|---|---|---|---|---|
| 50 | 6,688 | 953,208 | 341 MB | 10 MB | 0.7649 | 0.7098 | 0.7111 | 7.20% | 7.03% |
| 40 | 9,095 | 1,090,465 | 631 MB | 15 MB | 0.7840 | 0.7281 | 0.7294 | 8.06% | 6.96% |
| 30 | 13,266 | 1,268,143 | 1,343 MB | 21 MB | 0.8243 | 0.7561 | 0.7731 | 8.27% | 6.20% |
| 20 | 21,757 | 1,529,814 | 3,612 MB | 35 MB | 0.8660 | 0.8253 | 0.8311 | 4.70% | 4.04% |
| 10 | 43,057 | 1,914,590 | 14,144 MB | 67 MB | 0.9193 | * | 0.8615 | * | 6.29% |
| 5 | 61,959 | 2,110,399 | 29,289 MB | 100 MB | 0.9180 | * | 0.8492 | * | 7.49% |

### 5.6 Fast-GCCA-ISSM Experiment

As mentioned in section 3, high memory requirement for storing and processing matrices is a critical problem in generalized canonical correlation analysis, which limited the scalability of the proposed algorithm. The Fast-GCCA-ISSM is a fast and scalable estimation method which reduces memory requirement of GCCA-ISSM. In order to investigate performance of Fast-GCCA-ISSM and compare it with GCCA, we considered "Book" as the auxiliary and "Kindle" as the target domain. We designed this experiment based on the minimum number of common user's ratings in auxiliary and target domain. We reduced this parameter from 50 to 5 and applied, MF-GCCA-ISSM and MF-Fast-GCCA-ISSM for each setting. The experiments results are presented in Table 6. The symbol '*' in the last two rows of the table means that for the last two experiments we could not apply MF-GCCA-ISSM due to lack of required memory. It can be seen in all the experiments, accuracy of MF-Fast-GCCA-ISSM reduced compared to MF-GCCA-ISSM. But the reduction could be ignored against the reduction of required memory on large datasets. It could be concluded the scalability of MF-Fast-GCCA-ISSM is acceptable and it could be used in large cross-domain recommendation problems.

## 6 CONCLUSION

In this paper we presented a novel cross-domain recommendation algorithm using generalized canonical correlation analysis. First, we propose to use GCCA to solve the cold-start problem for new users in recommender systems. Second, we introduced an enhanced version of GCCA called GCCA-ISSM by inversing the sum of selection matrices that improves learning domain independent features for cold-start users. Using simulation experiments we demonstrated that GCCA-ISSM has better accuracy than GCCA in reconstructing less observed samples in cross-domain scenarios. Finally, we proposed an iterative algorithm called MF-GCCA-ISSM and MF-Fast-GCCA-ISSM for cross-domain recommender system. We performed various experiments on two real recommender system datasets. The experiments shows that: (1) The proposed method is better than the state of the arts baseline in cross-domain ratings prediction. (2) The proposed method always obtains better results than single domain MF specially for cold-start users. This improvement over single domain MF depends on various parameters such as number of common users among domains, auxiliary and target domains density, the correlation between domains and etc. (3) The proposed method could be extended by using multiple auxiliary domains to improve the recommendation accuracy. (4) Various auxiliary information could be used in the proposed method such as user reviews, items description, users demographic information and etc.

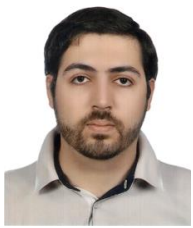

**Seyed Mohammad Hashemi** received the B.S. in Software engineering from Amirkabir University of Technology (Tehran Polytechnic), Iran, in 2010 and the M.S. in Artificial intelligence from Amirkabir University of Technology, Iran, in 2013. He is currently Ph.D. candidate in artificial intelligence in computer engineering department, Amirkabir University of Technology. His research interest includes machine learning, pattern recognition and data mining.

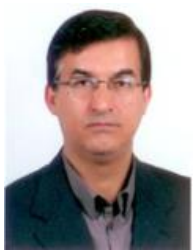

**Mohammad Rahmati** received the MSc in electrical engineering from the University of New Orleans, USA in 1987 and the PhD degree in electrical and computer engineering from University of Kentucky, Lexington, KY USA in 1994. He is currently a professor at the Computer Engineering Department, Amirkabir University of Technology (Tehran Polytechnic). His research interests are in the fields of machine learning, pattern recognition, image processing and video processing. He is a member of IEEE Signal Processing Society.